# Convolutional Neural Network: Text Classification Model for Open Domain Question Answering System

Muhammad Zain Amin[1,] Noman Nadeem[2]
Department of Computer Science and Engineering,
[1,2]University of Engineering and Technology, Lahore, Pakistan

## Abstract

Recently machine learning is being applied to almost every data domain one of which is Question Answering Systems (QAS). A typical Question Answering System is fairly an information retrieval system, which matches documents or text and retrieve the most accurate one. The idea of open domain question answering system put forth, involves convolutional neural network text classifiers. The Classification model presented in this paper is multi-class text classifier. The neural network classifier can be trained on large dataset. We report series of experiments conducted on Convolution Neural Network (CNN) by training it on two different datasets. Neural network model is trained on top of word embedding. Softmax layer is applied to calculate loss and mapping of semantically related words. Gathered results can help justify the fact that proposed hypothetical QAS is feasible. We further propose a method to integrate Convolutional Neural Network Classifier to an open domain question answering system. The idea of Open domain will be further explained, but the generality of it indicates to the system of domain specific trainable models, thus making it an open domain.

## Keywords

Machine Learning; Information Retrieval System; Convolutional Neural Network

## 1. Introduction

Deep learning models have achieved remarkable results in computer vision and speech recognition in recent years. Within natural language processing, much of the work with deep learning methods has involved learning word vector representations through neural language models and performing composition over the learned word vectors for classification. Word vectors, wherein words are projected from a sparse, 1-of-V encoding (here V is the vocabulary size) onto a lower dimensional vector space via a hidden layer, are essentially feature extractors that encode semantic features of words in their dimensions.

In Question Answering Systems deep learning is implemented in many different ways. The main purpose of a QA system is to find out ''WHO did WHAT to WHOM, WHERE, WHEN, HOW and WHY?'' [2].Some of the techniques being implemented are document Clustering, question classification and Named-Entity recognition. With the advent of new Amazon Alexa voice services NLP has gained so much focus either in speech recognition or in question answering system. Machine learning particularly Neural Networks has shown the promising results in these domains, which makes them more reliable.

Question answering is versatile which involves natural language processing, artificial intelligence, information technology and many more. Question classification, information retrieval and answer extraction are the three general components for question answering system [1] . Question classification is grouping the question according to the type of its entity. Information retrieval method is to extract information from documents, exploring for documents themselves and searching web database such as the internet, for text, images or data [6] [7] [8]. Answer extraction is a final stage in question answering system, in which we get our final answer of the given question.

Convolution Neural Network (ConvNets) involves a series of filters of different sizes and shapes which convolve (roll over) the original sentence matrix to reduce it into further low dimension matrices. In text classification ConvNets are being applied to distributed and discrete word embedding [3] [4] [5] [19]. The down sampling technique used in convolutional neural network is L2 Regularization. CNN utilizes an activation function which helps it run in kernel (i.e) high dimensional space for neural processing.

For Natural language processing, text classification is a topic in which one needs to set predefined classes to free-text documents. Text categorization is the research focus and key technology in the field of information retrieval and data mining since the amount of electronic text information has been rapidly increasing [9].

Text classification mainly focus on three topics which includes: Feature Engineering: most used feature is the bag-of-words feature, some more multiplex feature are designed such as part-of-speech tags, tree kernels and noun phrases, [10] [11] Feature Selection: goals at deleting noisy features and refine the classification performance, most common feature selection method is removing the stop words (e.g., "the") [12], Machine Learning Algorithms: use classifiers such as naïve Bayes (NB), logistic Regression (LR) and support vector machines (SVM).

In the present work we train the CNN model on consumer complaints dataset. Training dataset has more than 5 lac rows and 11 categories. CNN model is trained with 2 layers one is convolution and other is softmax layer. These two layers are on top of word embedding layer. With little tuning of parameters we achieve significant change in overall accuracy of model.

## 2. Related work

Sanjay K Dwivedi and Vaishali Singh provided an overview of the approaches of the question-answer system. Question answer system (QA) have three stages, i.e *Question analysis*: parsing, question classification, and query reformulation; *Document analysis*: extract possible documents and recognize the answer, and *Answer analysis*: extract possible answer and grade the best one. They discussed classification for characterizing question answer approaches which includes:

*Linguistic Approach*: depending on the structure and knowledge of the language, fetched same meaning in the different expression. Linguistic techniques such as tokenization, POS tagging, and parsing are executed to user's question and extract the particular response from the structured database.

*Statistical Approach*: is used for a large amount of data having sufficient amount of data for statistical comparison to be considered important. This approach is made to provide the enough amount of learning data during training statistical models, and the system could possibly deliver the positive response of even complex questions if the statistical model has been properly trained.

*Pattern Based Approach*: is an astonishing efficient technique for utilizing the web as a source data. Instead of linguistic analysis, pattern-based approach utilizes eloquence of text pattern, pattern matching not only decreasing the linguistic computations but also help in automatic wrapper generation for controlling the divergent web data [14].

Some of the previous literature review such as survey of text question answering techniques explained the types of Question Answering system which includes: Web based Question Answering system (using search engines to get back hyperlink that containing answers to the questions), IR/IE Question Answering system (returning a group of top ranked documents as responses to the query), Restricted Domain Question Answering system (needed a linguistic support to identify the natural language text to get the answer of the questions correctly), Rule Based Question Answering system (extended form for IR based question answering system) and Classification of Questioners level (the questions are classified into different levels according to its context) in the paper [1].

Xiang Zhang, Junbo Zhao and Yann LeCun provided an overview on character-level Convolutional Networks for text classification. They offer an actual method on the use of character-level convolutional networks for text classification. Applying convolutional network to text classification shown that convolutional network can be directly applied to distinct set of words without any information on the syntactic or semantic structures of the languages. They also constructed several large scale datasets to show that character-level convolutional network could achieve competitive results [15].

Siwei Lai, Liheng Xu, Xang Liu, Jun Zhao provided an overview for text classification through recurrent convolutional neural network without human desired features. In this paper, they provide the model of deep neural network to learn the semantics of the text and using the four datasets such as Fudan set (Chinese document classification that consist of twenty categories includes energy, education and art), 20Newsgroups (Choose four major headings like comp, politics, rec, religion from twenty newspaper), Stanford Sentiment Treebank (dataset which contains information about movie reviews and classified in different ways like Very Negative, Negative, Neutral, Positive, Very Positive), ACL Anthology Network (dataset holds the scientific documents) which are initially trained through different learning methods such as word representation learning (by combination of a word and its context we get a more precise word meaning) and text representation learning (represents the text data). They also use the max-pooling layer in text representation learning for down-sampling the data that automatically judges the most important semantic factors in the dataset and determine which word play key role in text classification to get the key components in texts [12] [16].

Kunlun Li, Jing Xie, Xue Sun, Yinghui Ma and Hui Bai provided an overview on multi-class text categorization based on LDA (Latent Dirichlet Allocation) and SVM (Support Vector Machine). Introducing LDA model into feature selection and merge into the relations between feature words and classes and to organize their generative model according to different types of textual data. To build and employing LDA model is to obtain the composition information of hidden topic. Multi-class SVM classifier is trained based on the implicit text matrix which is acquired from the feature words. The combination of the text representation with the powerful classification ability results confirms the effectiveness and superiority of this method [18].

## 3. Model

The implemented Convolution Neural Network model has three layers. Basic functionality of convolution neural network resembles to the visual cortex of the animal brain. In text classification task convolution neural network gives promising result. Criteria for text classification is similar to image classification only difference is that instead of pixel values we have matrix of word vectors. Proposed model is implemented in python using tensorflow library.

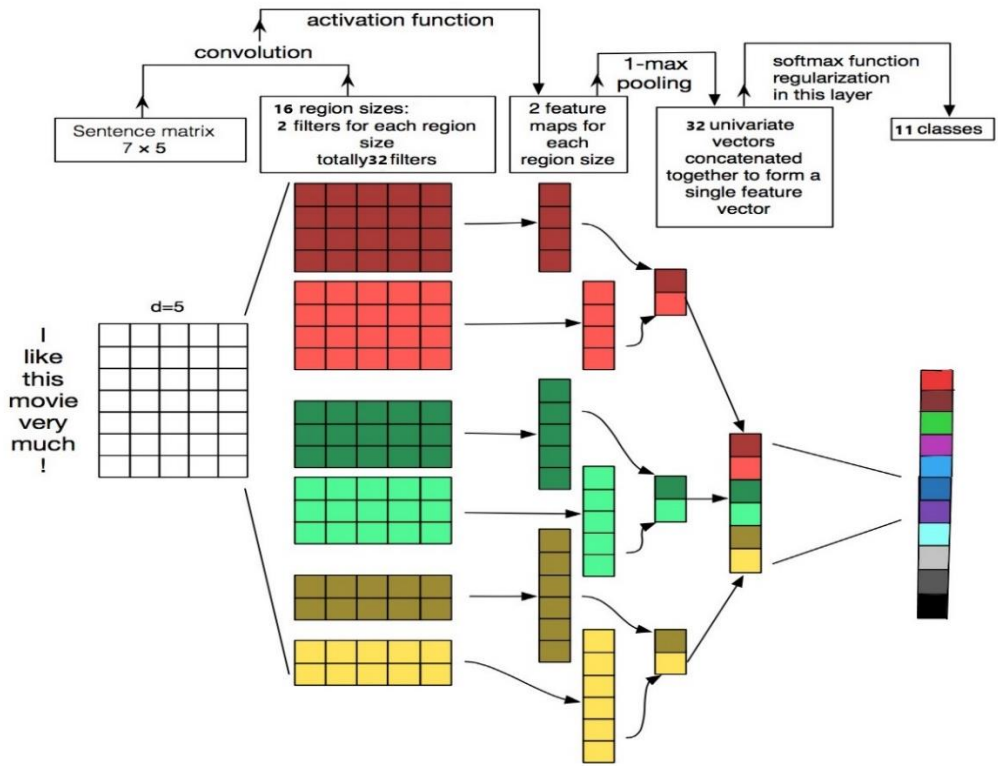

Figure 1: Convolutional Neural Network Model for 11 classes. Model is presented by using example sentence model is similar to model presented in this paper [17].

*3.1. Target Function*

Target Function employed includes neurons with learnable weights and biases. Each neuron receives several input, take weighted sum over them and pass it through an activation function and responds with an output. Whole network has a loss function calculated by spreading output to softmax layer. Softmax is a fully connected layer which also performs function of down sampling for the output.

*3.2. Representation*

The first layer in convolution neural network is embedding layer which maps vocabulary word indices to low dimensional vectors. It is essentially a lookup table that we learn from data. Vocabulary size is decided by getting the maximum sentence length $V^{wrd}$. Given a sentence of *N* words *W* every word is transformed into its respective embedding. After transformation of all the words to vectors these are than fed to the convolution layer.

*3.3. System Structure*

The model implemented includes three layers. First is the Embedding layer which convert the words to its respective embedding vectors, second is the convolution layer in which main processing of the model takes place. Predefined filters roll over the sentence matrix and reduce it into low dimensional matrix. Third layer is softmax layer which is a down sampling layer capable of reducing sentence matrix and calculating loss function.

Embedding lookup function is used to get the word embedding of the sentence. The matrix generated as a result of embedding layer is than padded to equalize all the sentences. The defined filters will than start reducing the matrix and generate convolved features. These convolved features are than further reduced. The output generated as a result of convolved features is than spread over the max pooling layer for further down sampling of output.

Filters of different sizes and shapes are defined. The shape of the filters use in the proposed model is (3, 4, 5). Total number of filter is 33. The filters will roll over the original sentence matrix thus reducing it to low dimensional matrix. Instead of training our own embedding we use embedding lookup function of tensorflow. Embedded sentences are than padded to make all the sentence matrices of same size and shape.

## 4. EXPERIMENTAL EVALUATION

We conduct series of experiment on convolution neural network using dataset described as follow:

*4.1. Dataset*

The dataset used is of Consumer Complaints [12]. There are total 11 categories in which data is classified. Total number of rows is 555958. Dataset is in csv MS Excel format. Dataset is split into train, test and validation set. We used predefined function of split dataset.

*4.2. Learning*

We experimented with model to check the flexibility of it. List of parameters is as follow:

Table 1. Default Parameters.

| Parameter Name | Value |
|---|---|
| Number of epochs | 1 |
| Batch size | 37 |
| Number of filters | 32 |
| Filter sizes | (3,4,5) |
| Embedding dimension | 50 |
| L2 Regularization | 0.1 |
| Evaluate every | 200 |
| Dropout probability | 0.5 |

We experimented with parameters by increasing number of epochs we found significant change in the results. Batch size and number of filters are defined according to the dataset. These are the default for this dataset. Embedding dimensions is defined according to the maximum sentence length. Evaluation of training takes place every 200 steps on the validation set. Dropout probability is used to down sample the output by 0.5 %.

*4.3. Evaluation Criteria*

An evaluation criterion is based on the following metrics:

*4.3.1. Accuracy*

Accuracy is a measure of percentage value for correct predictions of data. It is calculated by the following formula:

$$Accuracy = \frac{correct\ predictions}{all\ predictions}$$

*4.3.2. Precision*

Precision is a measure for which model is correctly predicting any particular category. It is calculated by following formula:

$$Precision = \frac{particular\ category\ predicted\ correctly}{all\ category\ predictions}$$

*4.3.3. Recall*

Recall is calculated by following formula:

$$Recall = \frac{correctly\ predicted\ category}{all\ real\ categories}$$

*4.3.4. F1 Score*

F1 score is a harmonic mean of precision and recall.

*4.3.5. Confusion Matrix*

It is a matrix which maps the predicted outputs across actual outputs.

### 4.4. Experiment 1

Parameters for first experiment are as follow:

Table 2. Parameters for Experiment 1.

| Parameter Name | Value |
|---|---|
| Number of epochs | 1 |
| Batch size | 37 |
| Number of filters | 32 |
| Filter sizes | (3,4,5) |
| Embedding dimension | 50 |
| L2 Regularization | 0.1 |
| Evaluate every | 200 |
| Dropout probability | 0.5 |

The accuracy achieved by using these parameters is 79%.

Table 3. Experiment 1 results

| Accuracy (%) | Precision (%) | Recall (%) | F1-Score (%) |
|---|---|---|---|
| 79.6 | 73.35 | 56.69 | 59.98 |

Table 4. Confusion matrix experiment 1

| Classes | 1 | 2 | 3 | 4 | 5 | 6 | 7 | 8 | 9 | 10 | 11 |
|---|---|---|---|---|---|---|---|---|---|---|---|
| [1] Customer Loan | 60 | 6 | 10 | 3 | 6 | 0 | 1 | 0 | 0 | 2 | 0 |
| [2] Bank account or service | 2 | 28 | 1 | 4 | 10 | 0 | 3 | 0 | 0 | 0 | 0 |
| [3] Credit card | 5 | 1 | 89 | 7 | 18 | 0 | 3 | 0 | 0 | 0 | 0 |
| [4] Credit reporting | 0 | 1 | 4 | 138 | 18 | 0 | 4 | 0 | 0 | 0 | 0 |
| [5] Debt collection | 2 | 3 | 6 | 17 | 242 | 0 | 6 | 0 | 0 | 1 | 5 |
| [6] Money transfers | 8 | 0 | 2 | 0 | 1 | 2 | 0 | 0 | 0 | 0 | 0 |
| [7] Mortgage | 1 | 3 | 0 | 2 | 4 | 0 | 205 | 0 | 0 | 0 | 0 |
| [8] Payday loan | 0 | 0 | 0 | 0 | 2 | 0 | 0 | 0 | 0 | 0 | 0 |
| [9] Prepaid card | 0 | 8 | 0 | 0 | 6 | 0 | 0 | 0 | 5 | 0 | 0 |
| [10] Virtual currency | 3 | 0 | 2 | 0 | 2 | 0 | 0 | 0 | 0 | 9 | 0 |
| [11] Student loan | 1 | 3 | 0 | 0 | 7 | 0 | 0 | 0 | 0 | 0 | 18 |

### 4.5. Experiment 2

We find that by increasing the number of epochs there is significant change in results. Parameters for second experiment are as follow:

Table 5. Parameters for Experiment 2.

| Parameter Name | Value |
|---|---|
| Number of epochs | 2 |
| Batch size | 37 |
| Number of filters | 32 |
| Filter sizes | (3,4,5) |
| Embedding dimension | 50 |
| L2 Regularization | 0.0 |
| Evaluate every | 400 |
| Dropout probability | 0.5 |

The accuracy achieved on second experiment is 80%.

Table 6. Experiment 2 results

| Accuracy (%) | Precision (%) | Recall (%) | F1-Score (%) |
|---|---|---|---|
| 80.3 | 75.25 | 55.66 | 59.30 |

Table 7. Confusion matrix experiment 2

| Classes | 1 | 2 | 3 | 4 | 5 | 6 | 7 | 8 | 9 | 10 | 11 |
|---|---|---|---|---|---|---|---|---|---|---|---|
| [1] Customer Loan | 66 | 0 | 8 | 4 | 8 | 0 | 1 | 0 | 0 | 1 | 0 |
| [2] Bank account or service | 1 | 28 | 3 | 2 | 12 | 0 | 2 | 0 | 0 | 0 | 0 |
| [3] Credit card | 7 | 1 | 92 | 5 | 17 | 0 | 1 | 0 | 0 | 0 | 0 |
| [4] Credit reporting | 1 | 0 | 4 | 138 | 18 | 0 | 3 | 0 | 0 | 0 | 0 |
| [5] Debt collection | 2 | 4 | 8 | 14 | 245 | 0 | 5 | 0 | 0 | 0 | 4 |
| [6] Money transfers | 8 | 0 | 1 | 0 | 1 | 3 | 0 | 0 | 0 | 0 | 0 |
| [7] Mortgage | 3 | 2 | 1 | 2 | 5 | 0 | 202 | 0 | 0 | 0 | 0 |
| [8] Payday loan | 0 | 0 | 0 | 0 | 2 | 0 | 0 | 0 | 0 | 0 | 0 |
| [9] Prepaid card | 0 | 7 | 0 | 0 | 9 | 0 | 0 | 0 | 3 | 0 | 0 |
| [10] Virtual currency | 5 | 0 | 4 | 0 | 2 | 0 | 0 | 0 | 0 | 5 | 0 |
| [11] Student loan | 1 | 4 | 0 | 0 | 4 | 0 | 0 | 0 | 0 | 0 | 20 |

*4.6. Experiment 3*

Parameters for third experiment are as follow:

Table 8. Parameters for Experiment 3.

| Parameter Name | Value |
|---|---|
| Number of epochs | 4 |
| Batch size | 37 |
| Number of filters | 32 |
| Filter sizes | (3,4,5) |
| Embedding dimension | 50 |
| L2 Regularization | 0.0 |
| Evaluate every | 400 |
| Dropout probability | 0.5 |

Accuracy achieved by using these parameters is 84%.

Table 9. Experiment 3 results

| Accuracy (%) | Precision (%) | Recall (%) | F1-Score (%) |
|---|---|---|---|
| 84.7 | 79.94 | 65.48 | 68.92 |

Table 10. Confusion matrix experiment 3

| Classes | 1 | 2 | 3 | 4 | 5 | 6 | 7 | 8 | 9 | 10 | 11 |
|---|---|---|---|---|---|---|---|---|---|---|---|
| [1] Customer Loan | 66 | 0 | 11 | 5 | 2 | 0 | 4 | 0 | 0 | 0 | 0 |
| [2] Bank account or service | 2 | 32 | 3 | 2 | 7 | 0 | 2 | 0 | 0 | 0 | 0 |
| [3] Credit card | 3 | 2 | 98 | 8 | 11 | 0 | 1 | 0 | 0 | 0 | 0 |
| [4] Credit reporting | 0 | 1 | 2 | 148 | 12 | 0 | 2 | 0 | 0 | 0 | 0 |
| [5] Debt collection | 1 | 1 | 7 | 17 | 247 | 0 | 4 | 0 | 0 | 0 | 5 |
| [6] Money transfers | 7 | 0 | 2 | 0 | 0 | 4 | 0 | 0 | 0 | 0 | 0 |
| [7] Mortgage | 0 | 1 | 1 | 1 | 3 | 0 | 209 | 0 | 0 | 0 | 0 |
| [8] Payday loan | 0 | 0 | 0 | 0 | 2 | 0 | 0 | 0 | 0 | 0 | 0 |
| [9] Prepaid card | 1 | 4 | 0 | 0 | 8 | 0 | 1 | 0 | 5 | 0 | 0 |
| [10] Virtual currency | 0 | 0 | 2 | 0 | 1 | 0 | 0 | 0 | 0 | 13 | 0 |
| [11] Student loan | 1 | 0 | 0 | 1 | 2 | 0 | 0 | 0 | 0 | 0 | 25 |

*4.6. Evaluation Results of Model*

Table 11. Final Results.

| Experiment # | Accuracy | Precision | Recall | F1 Score |
|---|---|---|---|---|
| 1 | 79.6 | 73.35 | 56.69 | 59.98 |
| 2 | 80.3 | 79.25 | 56.66 | 59.30 |
| 3 | 84.7 | 79.94 | 65.48 | 69.92 |

## 5. Integration with question answering system

We devise a model for open domain question answering system (QAS). System is composed of domain specific trainable classification models. For every data domain classification model is trained on the data of some specific domain. The idea is based on Quran question answering system which retrieve the most accurate verse from Quran related to the user question. The system works on neural network classifiers which classify the verses based on their contents and thus retrieving the most related verse, but the difference with our system is that it can be trained on any number of classifiers for certain set of domains which makes it an open domain platform for question answering system.

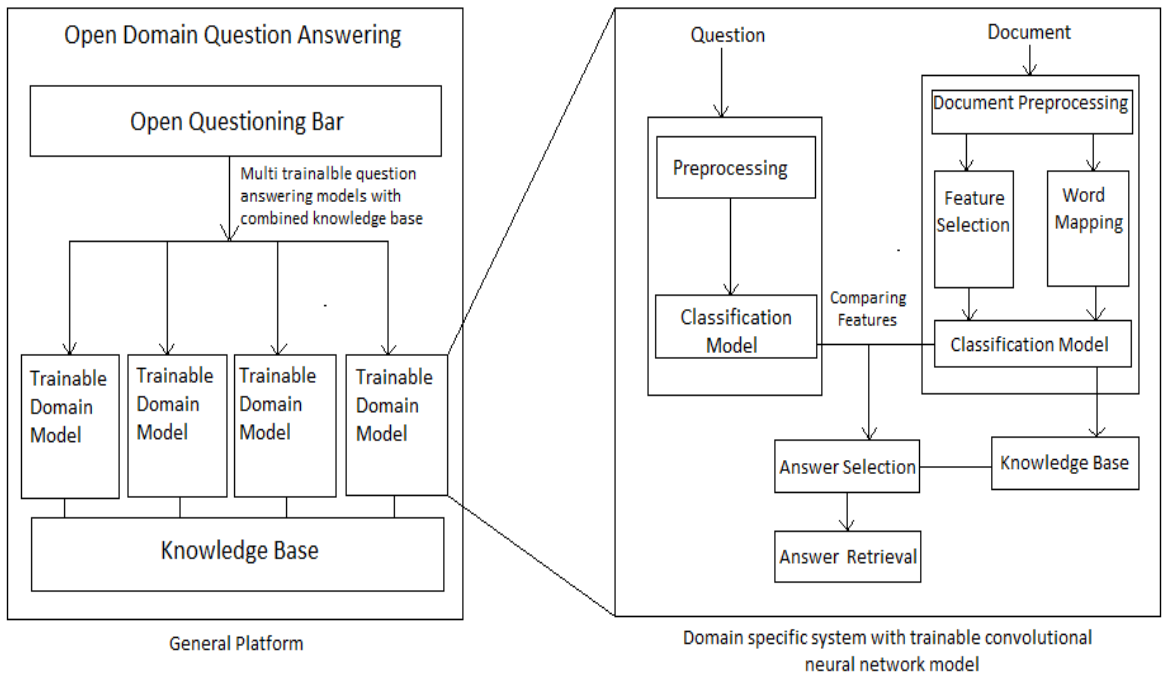

Figure 2: Proposed Model for Integration with Question Answering System

The proposed QAS has four basic modules.

*5.1. General Platform*

Open Domain Question Answering System includes number of trainable models. Model can be trained on some specific dataset. All the models will have a combined knowledgebase. General question answering bar will answer questions by using combined knowledgebase. Performance of the system depends on the classification models.

*5.2. Document Processing*

Document Processing is an offline phase in which user will enter the document related to any specific domain. By using Natural Language Processing techniques features are selected from the text. These features will be than use to categorize the text. Feature selection is an important phase because the categories defined in this phase will be used for the classification purpose. Proposed QAS also employ word mappings. Mappings are used to define the word embedding. Obtained categories of text along with related text and their respective word embedding are used to classify text using proposed Convolution Neural Network Model. Neural Network Model will develop its knowledge base which is not more than the vocabulary of the processed document.

*5.3. Question Processing*

Question processing can be online. Once the classification model is trained on the data, the same model can be used to classify the questions. Question is tokenized and tagged using natural language processing techniques. Question will be classified into categories obtained as a result of document processing. Once the successful classification of question occurs system will compare the semantic information of related category text to the question.

*5.4. Answer Retrieval*

By comparing semantic information of question to related category text in which question is classified answer selection occurs. System will also add the new semantic information to knowledge base so that it can learn and handle diverse question.

## 6. Discussion and Future Work

Question answering system is fairly an information retrieval system. Over past few years there has been significant development in the field of question answering. But most of the systems developed are domain specific some can.  We devise the method to develop an open domain question answering system by using trainable classification models. Our classification model is flexible enough to handle different kinds of dataset. This will help justify the fact that proposed question answering system is feasible. Classification model can handle large scale dataset. Question answering system will employ the number of trainable classification models for, or QAS to be open domain it must have a general knowledgebase so it can add semantic information from different domains.

The future work will be to establish an automatic feature selection mechanism which will extract features from the document text and model the document based on those features. Further techniques which question answering system will employ are relation extraction, knowledge base and information retrieval.

# 7. Conclusion

The paper provides the implementation of a convolution neural network text classifier. ConvNets has shown promising results on large scale text classification. Furthermore we presented an idea of open domain question answering system employing trainable classification models. Also describes the method to integrate the convolution neural network model with the proposed open domain question answering system.